# Functional Nanoscale Phase Separation and Intertwined Order in Quantum Complex Materials

Gaetano Campi [1*], Antonio Bianconi [1,2,3]

[1] Institute of Crystallography, CNR, Via Salaria Km 29.300, Monterotondo Rome, 00015 Italy; gaeta-no.campi@ic.cnr.it
[2] RICMASS Rome International Center for Materials Science, Superstripes Via dei Sabelli 119A, 00185 Roma, Italy; antonio.bianconi@ricmass.eu
[3] National Research Nuclear University MEPhI (Moscow Engineering Physics Institute), 115409 Moscow, Russia

**Abstract**:
Nanoscale phase separation (NPS) characterized by particular types of correlated disorder plays an important role in the functionality of high temperature superconductors (HTS). Our results show that multiscale heterogeneity is an essential ingredient of quantum functionality in complex materials. Here, the interactions developing between different structural units cause dynamical spatiotemporal conformations with correlated disorder. Thus visualizing conformations landscape is needed for understanding the physical properties of complex matter and requires advanced methodologies based on high precision X ray measurements. We discuss the dynamical correlated disorder at nanoscale and the related functionality in oxygen doped perovskite superconducting materials.

## 1. Introduction

Nanoscale phase separation (NPS) has been considered to be detrimental for high temperature superconductivity. However, high precision X-ray measurements in the solid-state physics [1-4] have recently provided experimental validation for the alternative paradigm where lattice heterogeneity from the atomic limit to the micron scale plays a key role [6-12]. NPS shows intertwined and interlocked nanoscale lattice structures forming heterostructures at atomic limit [13]. This induce the emergence of novel functionalities such as high temperature superconductivity in complex quantum materials like organics [13-20], doped perovskites [21-39], iron based superconductors [40-44], charge density wave materials [45-52], manganites showing colossal magnetoresistance [53-57], doped diborides [58-60], smart structures and biological systems [61-66], perovskite oxide interfaces [67-69] and magnetic materials [70-71].
Today, at the beginning of the twenties of the XXI century, the development of advanced experimental X-ray methods has provided evidence that the quantum functionality, e.g. colossal magnetoresistance and high temperature superconductivity, are controlled by the correlated lattice disorder from atomic scale to micron scales. This multiscale lattice heterogeneity has been quite difficult to be unveiled by conventional experimental approaches, requiring high spatial resolution probes. Nowadays, advanced latest generation synchrotron sources and new X-ray optics, have allowed the design of new scanning techniques based on focused X ray beams.

Several crystallographic structures coexisting on micron- and nano-scale have been visualized in high temperature superconductors (HTS). Strain and doping drive these systems at critical points for phase separation with coexistence of different crystallographic phases in different nano-regions of the crystal. While in the literature of HTS, the focus of theoretical research has been addressed to electronic phase separation and inhomogeneity due to competition or intertwining between superconductivity and magnetism, the main result of Scanning X Ray Diffraction has been to provide compelling experimental evidence that at nanoscale different puddles show a substantial lattice structural difference, which necessarily implies electronic and magnetic phase separation. In fact, looking at the nanoscale we have observed different nano-regions with different lattice structures showing different electronic and magnetic ordered phases with different topologies that influence the Fermi surfaces, the pseudogap energy and even the superconducting critical temperature. We would like to underline archetypal cases of intrinsic phase separation in $A_xFe_{2-y}Se_2$ originated by the coexistence of an insulating magnetic phase and a paramagnetic metallic phase with an in-plane compressed and expanded lattice [40-44]. A second clear case of phase separation has been found in $YBa_2Cu_3O_{6+y}$, with y<0.5 where super-conductivity (y>0.33) percolates in a network of oxygen ordered nano-regions (y=0.5), interspersed with oxygen depleted domains (y=0) [37].

Here, we have chosen to discuss the relationship between phase separation at nanoscale and the emerging macroscopic properties in two high temperature supercon-ducting quantum materials such as $La_2CuO_{4.1}$ [38] and $HgBa_2CuO_{4.12}$ [7] doped by mobile oxygen interstitial ions (O-i).

## 2. Perovskites doped by mobile Oxygen Interstitials

### *2.1. Correlated disorder and phase separation in $La_2CuO_{4+y}$*

$La_2CuO_{4+y}$ is one of the simplest compound where y oxygen interstitial ions (O-i) in the rocksalt $[La_2O_{2+y}]$ are intercalated by $[CuO_2]$ layers. Thanks to the high synchrotron photon flux, we have measured the satellite peaks, associated with super-cells due to oxygen interstitial O-i dopants and Charge Density Waves, CDW, arrangement (see Figure 1a). CDW incommensurate diffuse satellites with wave-vector $q_{CDW}$ = 0.21$b^*$ + 0.29$c^*$ coexist with O-i satellites displaced by $q_{O-i}$ = 0.25$b^*$ + 0.5$c^*$. O-i and CDW super-structures are characterized by a different staging, that is their $c^*$ component; indeed, O-i and CDW have staging $c^*$=0.5 and $c^*$=0.29, respectively, as indicated by the white arrows in panel (a).

A common and intriguing feature of quantum materials is that their emerging properties can be easily manipulated by weak external stimuli such as temperature gradient, strain, photon illumination. Due to the mobility of oxygen interstitials, O-i, our systems present structural conformations landscape. The conformations have been manipulated by continuous X ray photo illumination, as shown in Figure 2(b) where we bring the sample from a structural conformation with disordered O-i, to a different conformation with ordered O-i, after illumination. Different conformations characterized by different order degree of O-i [73,74] can be obtained by combining X ray flux intensity with thermal cycling, as shown in Figure 2(c).

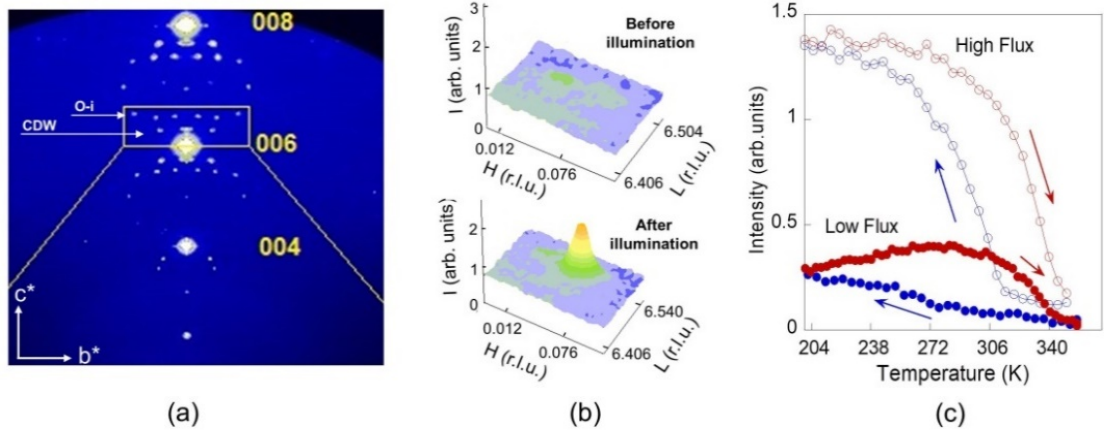

**Figure 1**: (a) XRD image of b*c* plane in the reciprocal lattice of the $La_2CuO_{4.1}$ sample. We observe several superstructures surrounding the indicated Bragg peaks (004), (006), (008) due to O-i ordering and Charge Density Waves (CDW). The arrows indicate the O-i and CDW superlattices. (b) At T= 330 K the O-i get disordered in the sample and the diffraction satellite vanishes (upper panel). X-ray photon illumination O-i allows speed up the ordering kinetic (lower panel). (c) The oxygen ordering rate depends on the X ray flux, as shown in the thermal cycle measured by using two different X ray photon flux [74]. Red and blue symbols refer to heating and cooling cycles, respectively.

Here the (full symbols) X ray low-flux and (empty symbols) high-flux on the sample correspond to $0.5 \cdot 10^{14}$ photons $s^{-1}$ $cm^{-2}$ and $5.0 \cdot 10^{14}$ photons $s^{-1}$ $cm^{-2}$, respectively. The temperature change rate has been of 1K min-1 in both low flux and high flux illumination. Each point on the two hysteresis corresponds with a specific structural conformation with specific degree of order (or disorder) of O-i.

The different O-i ordered domains are inhomogeneously distributed in space as seen by SµXRD. The maps in Figure 2 (a) and 2 (b) show two different O-i distributions obtained with two different thermal treatments; the first one (a) shows the maximum critical temperature ($T_c$=40K) and the second (b) presents a macroscopic phase separa-tion with two critical temperatures, $T_c$ = 16 K and 32 K [6].

We have characterized this inhomogeneity by using spatial statistics tools de-scribed in [6]. In both samples the probability density function of O-i satellite intensity follows an exponentially truncated power-law distribution, given by $P(x) = x^{-\alpha} \exp(-x/x_0)$ where α is the critical exponent and $x_0$ is the cut-off. The critical exponent, α, results to be 2.6, both in the high-$T_c$ (a) and low-$T_c$ (b) conformations indicating the fractal nature of O-i arrangement. On the other hand, the cut-off, $x_0$, is quite larger for the high-$T_c$=40 K sample as can be seen in Figure 2(c). This means that larger extent of O-i satellite intensity with fractal geometry favors a higher $T_c$ [38].

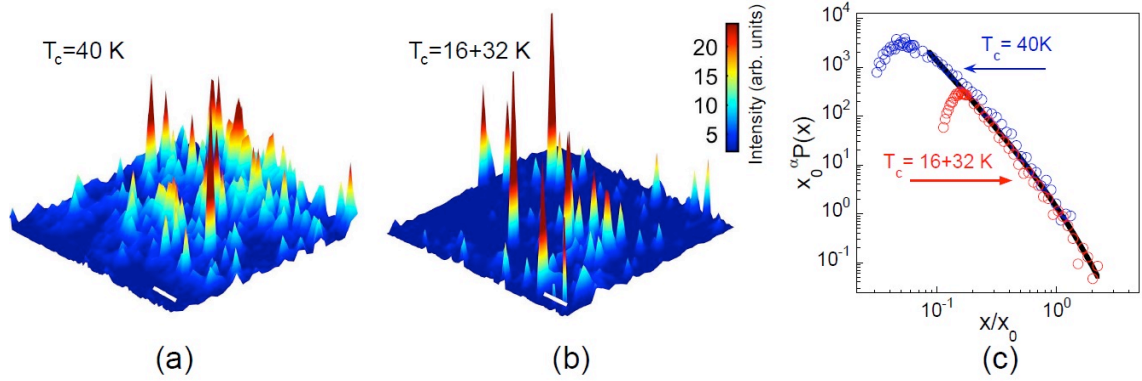

(a)     (b)     (c)

**Figure 2**: Spatial distribution of the O-i superstructure intensity for two different conformations in $La_2CuO_{4.1}$ sample with (a) $T_c=40$ K (b) and $T_c=16$ K and 32 K phases in a 500×400 µm$^2$ area. The bar corresponds to 50 µm. (c) The rescaled probability distribution, $x^{\alpha}P(x)$, of the O-i superstructure intensity for the two samples. The cut-off, $x_0$, increases from 8±1 in the low-Tc conformation to 31±2 in the high-Tc conformation. The rescaled distributions collapse on the same universal curve (black solid line) [38].

We have found that the spatial inhomogeneity is characterized by a strong spatial anticorrelation between the O-i rich and CDW rich regions. In the oxygen rich regions, the system is in the overdoped metallic phase, while in the oxygen poor region in the underdoped phase we measure high density of CDW. Figure 3(a) shows the map of difference between the intensity of O-i and CDW diffraction satellites. The O-i are located in the 1/4,1/4,1/4 site, and form commensurate stripes in the $La_2O_2$ structure, intertwined with incommensurate lattice CDW domains [39]. Indeed, we observe micron size puddles of blue metallic oxygen rich regions separated by red CDW rich zones, by a filamentary percolating interface indicated by white space in Figure 3(b).

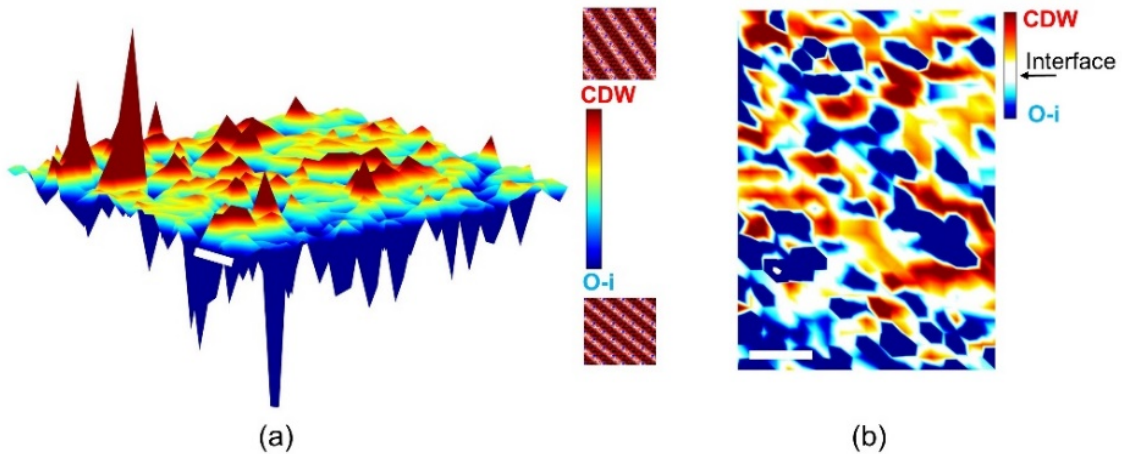

(a)     (b)

**Figure 3**: (a) Surface plot of the map difference between the normalized intensities of both CDW and O-i satellites in $La_2CuO_{4.1}$ [38]. CDW puddles are dominant in red regions while the O-i stripes are dominant in the blue zones. The lattice modulations corresponding to CDW and O-i super-structures are depicted. (b) 2D map difference highlighting the interface (white region) between CDW rich and O-i rich regions. The white bars correspond to 5 µm.

## 2.2. Correlated disorder and phase separation in $HgBa_2CuO_{4+y}$

Structural and electronic inhomogeneity has been studied in the high temperature perovskite superconductor with tetragonal crystal symmetry $HgBa_2CuO_{4+y}$ with a single $CuO_2$ plane. In this compound the oxygen interstitial ions (O-i), form atomic stripes in the spacer layer [$HgO_yBa_2O_2$] between [$CuO_2$] planes. We have measured the diffuse scattering associated with both CDW and oxygen interstitial arrangement in the lattice. Diffuse CDW satellites have been detected at $\mathbf{q_{CDW}} = 0.23\mathbf{a^*} + 0.16\mathbf{c^*}$ and $\mathbf{q_{CDW}} = 0.23\mathbf{b^*} + 0.16\mathbf{c^*}$ (where $\mathbf{a^*}$, $\mathbf{b^*}$ and $\mathbf{c^*}$ are the lattice units in the reciprocal space) around the (108) Bragg peak, below the onset temperature $T_{CDW}$ = 240 K [7]. Resolution-limited streaks connecting the Bragg peaks, due to atomic O-i stripes in the $HgO_y$ spacer layers have been measured in agreement with previous experiments [6,32].

In Figure 4(a) and 4(b) we show the maps of the integrated intensity of CDW peak and oxygen O-i diffuse streaks, respectively [7,9] as seen by SμXRD measurements. Both O-i and CDW maps are spatially inhomogeneous and their PDFs is well modeled by a power-law behavior also in this case, as shown in Figure 4(c). The critical exponents in the two intensity distributions are 1.8 ± 0.1 and 2.2 ± 0.1 for the O-i and CDW, respectively [7].

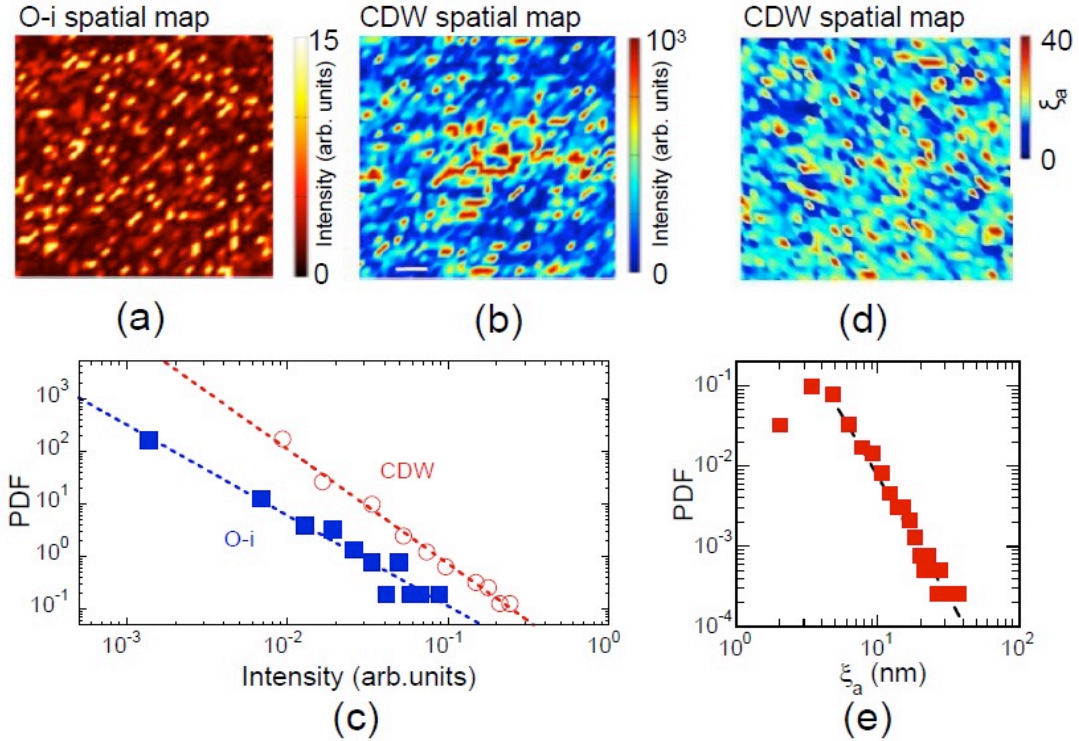

**Figure 4**: Colour plot of the map of the (a) O-i-streak intensity and (b) Charge Density Waves (CDW)-peak in $HgBa_2CuO_{4.12}$. (c) Probability density function calculated from the O-i streaks and CDW intensity map in (a) and (b), showing a power-law behaviour. (d) Colour map of the CDW coherence length $\xi_a$ of the CDW. (e) Probability density distribution of $\xi_a$.

A map of the spatial organization of the in-plane CDW-puddle size, $\xi_a$, is shown in Figure 4(d). Although the average size of CDW puddles is 4.3 nm (in agreement with previous works), its probability density function, shown in Figure 4(e), has a fat-tailed distribution fitted by a power-law curve with critical exponent of 2.8±0.1. Thus, rare and larger puddles up to 40 nm are measured and their distribution provide a complex topology for the flowing of superconductivity currents [7,9].

As in the $La_2CuO_{4+y}$ superconductor, the spatial inhomogeneity shows a negative correlation between O-i and CDW. This is well depicted in the 'difference map' between CDW peaks and O-i diffuse streaks in Figure 4(a). The poor CDW regions on the $CuO_2$ basal plane correspond to O-i rich regions on the $HgO_y$ layers. The O-i atomic striped domains are here intertwined with incommensurate lattice CDW. The perco-lating filamentary interface between O-i rich and O-i poor regions is indicated by white space in Figure 5(b)

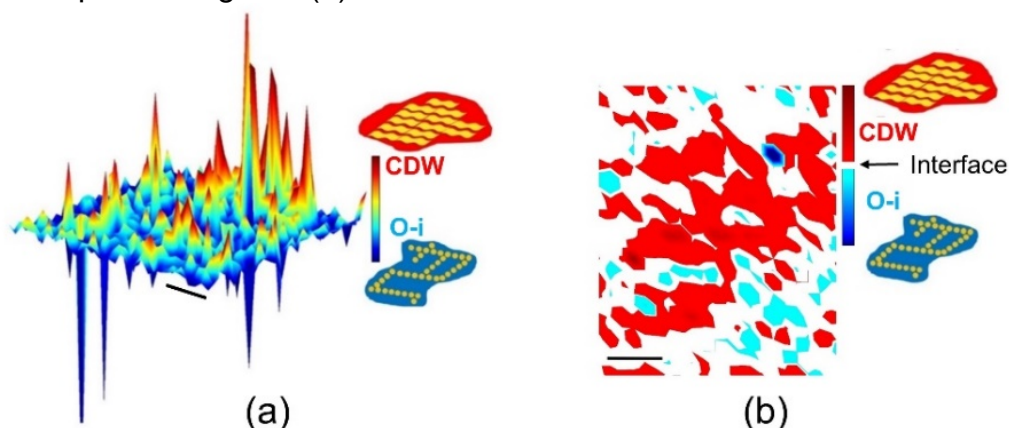

**Figure 5**: (a) Surface plot of the map difference between the normalized intensities of both CDW satellite and O-i streaks in $HgBa_2CuO_{4.12}$ [7]. CDW puddles are dominant in red regions and the O-i stripes are dominant in the blue zones. The bar corresponds to 5 μm. (b) Pictorial view of the spatial anticorrelation between CDW puddles in the CuO2 plane and the O-i rich domains in the $HgO_y$ layers. Here the interface space between CDW and O-i has been highlighted by the white region. The bar corresponds to 5 μm.

## 3. Discussion

The anomalous phase diagram of HTS cuprates is believed nowadays to be closely related to the phase separation at multiple scale length. This feature is particularly pronounced in those compounds doped by mobile oxygen interstitials. In this context, the present work reports evidence of nanoscale phase separation in the case of two different oxygen doped layered perovskites, $La_2CuO_{4+y}$ and $HgBa_2CuO_{4+y}$. At optimum doping level, specific charge ordering and segregation lead to the formation of metallic and insulating domains with oxygen-rich and oxygen-poor regions, respectively. These different zones have been measured as different superlattices in X Ray Diffraction and their inhomogeneous spatial distribution has been visualized by Scanning micro X Ray Diffraction. The experimental results provide relevant evidence for the universality of phase separation where dopants O-i rich domains are intertwined with CDW domains even in the most optimized superconducting cuprates.

We have shown that the coexisting phases can be easily manipulated, e.g. by X ray illumination on controlled small areas, drawing a way to extend the functionality of the investigated fractal materials in the direction of information storage [72-76]

## 4. Materials and Methods

The $La_2CuO_{4+y}$ (LCO) single crystal with y=0.1 has orthorhombic Fmmm space group symmetry with lattice parameters a=(5.386±0.004)Å, b=(5.345±0.008)Å, c=(13.205±0.031) Å at room temperature. The $HgBa_2CuO_{4+y}$ (Hg1201) single crystal with y=0.12 has a sharp superconducting transition at $T_c$=95 K. The crystal structure has tetragonal P4/mmm space group symmetry with lattice parameters

a=b=0.387480(5) nm and c=0.95078(2) nm at T=100K. Diffraction measurements on both single crystals of LCO and Hg1201, were performed on the ID13 beamline at ESRF as described in [38] and [7], respectively. Data analysis has been performed by using customized and homemade written MATLAB routines [6].

**Author Contributions**: G.C. and A.B. conceived the experiments and wrote the paper.
**Acknowledgments**: The authors thank Luisa Barba and XRD1 beamline staff at ELETTRA, Tri-este, Italy; Manfred Burghammer and ID13 beamline staff at ESRF; Alessandro Ricci, Nicola Poccia, Michela Fratini and Stefano Agrestini for the longstanding collaboration.
**Conflicts of Interest**: The authors declare no conflict of interest.